\documentclass[floats,aps,showpacs,twocolumn,superscriptaddress]{revtex4}

\usepackage[dvips]{graphicx}
\usepackage{amsfonts}
\usepackage{amsmath,amssymb}
\usepackage{dsfont}
\usepackage{placeins}
\usepackage{afterpage}
\usepackage{flafter} 

\newcommand{\heff}{h_{\text{eff}}}
\newcommand{\hbareff}{\hbar_{\text{eff}}}
\newcommand{\ud}{\text{d}}
\newcommand{\ue}{\text{e}}
\newcommand{\ui}{\text{i}}
\newcommand{\psireg}{\psi_{\text{reg}}}
\newcommand{\psich}{\psi_{\text{ch}}}
\newcommand{\Ureg}{U_{\text{reg}}}
\newcommand{\Uch}{U_{\text{ch}}}
\newcommand{\UV}{U_{\text{V}}}
\newcommand{\UT}{U_{\text{T}}}
\newcommand{\UVreg}{U_{\widetilde{V}}}
\newcommand{\UTreg}{U_{\widetilde{T}}}
\newcommand{\epsV}{\varepsilon_{V}}
\newcommand{\epsT}{\varepsilon_{T}}
\newcommand{\Hreg}{H_{\text{reg}}}

\begin{document}

\title{Regular-to-chaotic tunneling rates using a fictitious integrable system}
\author{A.~B\"acker, R.~Ketzmerick, S.~L\"ock, and L.~Schilling\\[0.1cm]
        \textit{\small Institut f\"ur Theoretische Physik, 
                       Technische Universit\"at
                       Dresden, 01062 Dresden, Germany}}
\date{\today}

\begin{abstract}
We derive a formula predicting dynamical tunneling rates from 
regular states to the chaotic sea in systems with a mixed phase space. 
Our approach is based on the introduction of a fictitious 
integrable system that resembles the regular dynamics within the island.
For the standard map and other kicked systems we find agreement with 
numerical results for all regular states in a regime where resonance-assisted 
tunneling is not relevant. 
\end{abstract}

\pacs{05.45.Mt, 03.65.Sq, 03.65.Xp}
\maketitle

Tunneling of a quantum particle 
is one of the central manifestations of quantum mechanics.
For simple 1D systems tunneling under a potential barrier 
is well understood and described, e.g.\
by using semiclassical WKB 
theory or the instanton approach \cite{WKBInstcomb}.
For higher-dimensional systems so-called ``dynamical tunneling'' 
\cite{DavHel1981} occurs
between regions which are separated by dynamically generated barriers.
Typically, such systems  have a mixed phase space 
in which regions of regular motion and irregular dynamics coexist.
Tunneling in these systems is barely understood as it generically cannot 
be reduced to the instanton or WKB approach. It has been studied theoretically 
\cite{HanOttAnt1984,Wil1986,BohTomUll1993,TomUll1994,Tom1998,Cre1998,FriDor1998,
BroSchUllEltComb,WimSchEltBuc2006,PodNar2003,SheFisGuaReb2006,
ShuIkeOniTakComb}
and experimentally, e.g. in cold atom systems
\cite{SteOskRai2001,HenHafBroHecHelMcKMilPhiRolRubUpc2001}
and semiconductor nanostructures \cite{FroWilHayEavSheMIuHen2002}.
A precise knowledge of tunneling rates is of current interest
for e.g.\ eigenstates affected by 
flooding of regular islands \cite{SchOttKetDit2001,BaeKetMon200507},
emission properties of optical microcavities \cite{WieHen2006} 
and spectral statistics in systems with a mixed phase space \cite{BerRob1984}.

There are different approaches for the prediction of tunneling rates
depending on the ratio of Planck's constant $h$ to
the size $A$ of the regular island.
In the semiclassical regime, $h\ll A$, 
small resonance chains inside the island dominate the tunneling
process (``resonance-assisted tunneling'')
\cite{BroSchUllEltComb,WimSchEltBuc2006}.
In contrast, we focus on the experimentally relevant regime 
of large $h$ (while still $h<A$), where small resonance chains 
are expected to have no influence on the tunneling rates.
This regime has been investigated in Ref.~\cite{PodNar2003},
however, the prediction does not seem to be generally applicable (see below).
Other studies in this regime investigate situations 
\cite{SheFisGuaReb2006,FeiBaeKetRotHucBur2006}, where dynamical tunneling 
can be described by 1D tunneling under a barrier, however, 
in our opinion they are non-generic. A generally applicable theoretical 
description of dynamical tunneling rates in systems with a mixed phase space 
is still an open question.

\begin{figure}[b]
  \begin{center}
    \vspace*{-0.25cm}
    \includegraphics[angle = 0, width = 85mm]{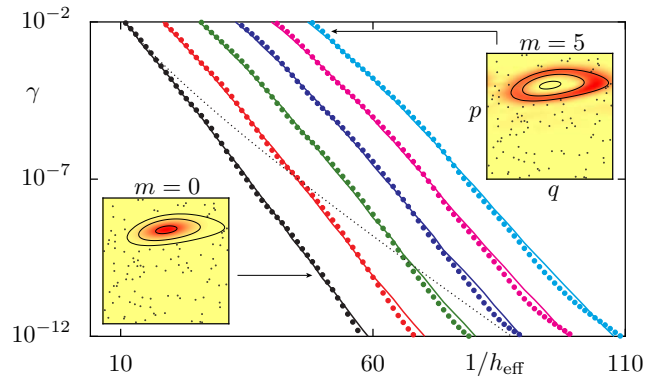}
    \vspace*{-0.4cm}
    \caption{(color online) Dynamical tunneling rates from a regular island 
          to the chaotic sea for the kicked system \cite{Map_amphib}: 
          Numerical results (dots) and prediction 
          following from Eq.~\eqref{eq:tunneling_rate_final} (lines) 
          vs $1/\hbareff$ for quantum numbers $m \leq 5$.
          The insets show Husimi representations of the regular states 
          $m=0$ and $m=5$ at $1/\heff=50$.
          The prediction of Ref.~\cite{PodNar2003} for $m=0$ with a fitted 
          prefactor is shown (dotted line).
         }
    \label{fig:distorted_island_amphib}
    \vspace*{-0.4cm}
  \end{center}
\end{figure}

\begin{figure}[t]
  \begin{center}
    \includegraphics[angle = 0, width = 85mm]
                     {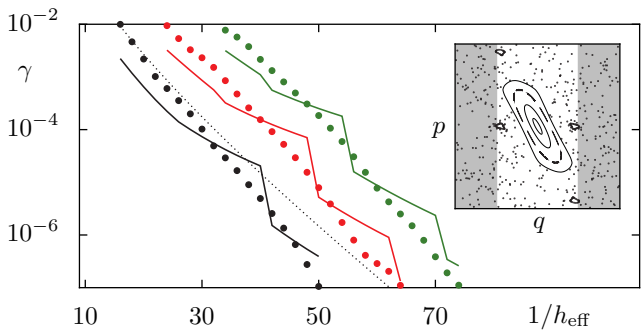}
    \vspace*{-0.4cm}
    \caption{(color online) Tunneling rates for the standard map ($K=2.9$) 
          for $m \leq 2$. 
          Prediction of Eq.~\eqref{eq:tunneling_rate_final} 
          (lines) and numerical results (dots), obtained using an absorbing 
          boundary at $q=\pm1/4$ (gray-shaded area of the inset). 
          Prediction of Ref.~\cite{PodNar2003} for $m=0$ with a fitted 
          prefactor (dotted line).
          }
    \label{fig:rates_standard_map}
    \vspace*{-0.6cm}
  \end{center}
\end{figure}

In this paper we present a new approach to dynamical tunneling from a regular 
island to the chaotic sea. The central idea 
is the use of a fictitious integrable system resembling the regular island. 
This leads to a tunneling formula involving properties of this integrable system 
as well as its difference to the mixed system under consideration. It allows 
for the prediction of tunneling rates from any quantized torus within the 
regular island. We find excellent agreement with numerical data, see
Fig.~\ref{fig:distorted_island_amphib}, for an example system 
where tunneling is not affected by phase-space 
structures like cantori at the border of the island. 
The applicability to more general systems is demonstrated for the standard map, 
see Fig.~\ref{fig:rates_standard_map}.

\begin{figure}[t]
  \begin{center}
    \includegraphics[]{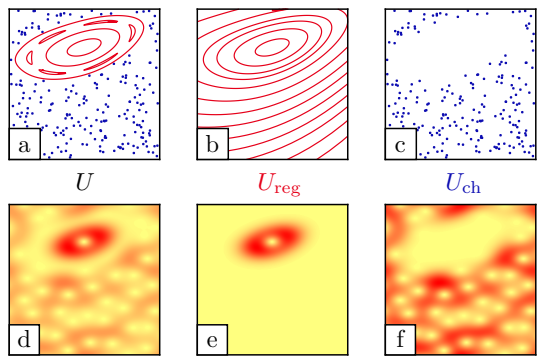}
    \vspace*{-0.2cm}
    \caption{(color online) (a-c) The classical phase 
          space corresponding to some quantum maps $U$, $\Ureg$, and $\Uch$. 
          (d-f) Husimi representation of eigenstates of such maps.
          Eigenstates of $U$ have a regular 
          and a chaotic component, as illustrated in the strongest form of a 
          hybrid state (d). Eigenstates $|\psireg\rangle$ 
          ($|\psich\rangle$) of $\Ureg$ ($\Uch$) are purely regular (chaotic).
          }
    \label{fig:quantum_maps}
    \vspace*{-0.65cm}
  \end{center}
\end{figure}

We consider 2D maps with one major regular island 
embedded in the chaotic region 
(Fig.~\ref{fig:distorted_island_amphib}, insets) which are described quantum 
mechanically by a unitary operator $U$ 
\cite{BerBalTabVor1979}. 
Classically the regular and chaotic region are separated, 
however, quantum mechanically they are coupled. 
This coupling has consequences for the eigenstates of $U$. While they are mainly
regular or chaotic, i.e. concentrated on a torus inside the regular region or 
spread out over the chaotic sea, they do have at least a small component in the 
other region. This is most clearly seen for hybrid states 
(Fig.~\ref{fig:quantum_maps}d). 
For a wave packet started on the $m$-th quantized torus 
($m=0,1,\dots,m_{\text{max}}-1$) coupled to an infinite chaotic sea the decay
$\ue^{-\gamma_m t}$ is described by a tunneling rate $\gamma_m$.
For systems with a finite phase space this exponential decay occurs at most up 
to the Heisenberg time $\tau_H=h/\Delta_{\text{ch}}$, where $\Delta_{\text{ch}}$ 
is the mean level spacing of the chaotic states. 
Introducing purely regular states 
$|\widetilde{\psi}_{\text{reg}}\rangle$ and orthogonal chaotic states 
$|\widetilde{\psi}_{\text{ch}}\rangle$ 
the tunneling rate
from such a purely regular state can be expressed by Fermi's golden rule
$\gamma = (2\pi/\hbar)\,\overline{|V|^2}\rho_{\text{ch}}$,
where $\rho_{\text{ch}}=1/\Delta_{\text{ch}}\propto N_{\text{ch}}$ 
is the chaotic density of states and $V=\langle\widetilde{\psi}_{\text{ch}}
\vert\widehat{H}\vert\widetilde{\psi}_{\text{reg}}\rangle$ 
for a time-independent $\widehat{H}$. For a map $U$
one replaces the local average over matrix elements in Fermi's golden rule by an
average over all $N_{\text{ch}}$ chaotic states and expresses $\gamma$
with respect to the time period of $U$, yielding
\begin{eqnarray}
  \label{eq:tunneling_rate_1}
    \gamma = \sum\limits_{\text{ch}} \left\vert v \right\vert^{2},
\end{eqnarray}
where $v=\langle\widetilde{\psi}_{\text{ch}}
\vert U\vert\widetilde{\psi}_{\text{reg}}\rangle$.
The eigenstates of $U$ cannot be used for determining the small matrix 
elements $v$, as they are neither purely regular nor purely chaotic. 

In order to construct purely regular and chaotic states we introduce
fictitious regular and chaotic quantum maps $\Ureg$ and $\Uch$ \cite{Rel_idea}. 
Here $\Ureg$ is 
regular in the sense that it can be written as 
$\ue^{-\ui\widehat{H}_{\text{reg}}/\hbareff}$, where $\Hreg$ is a 1D
Hamiltonian, which is integrable by definition and $\hbareff$ is the effective
Planck constant. 
$\Hreg$ has to be chosen such that its dynamics over one time unit resembles 
the classical motion corresponding to $U$ within the regular island as closely 
as possible (Fig.~\ref{fig:quantum_maps}b). 
The eigenstates $|\psireg\rangle$ of $\Ureg$ are localized in the regular 
region and continue to decay into the chaotic sea 
(Fig.~\ref{fig:quantum_maps}e). This is the decisive property of
$|\psireg\rangle$, which is in contrast to those eigenstates of $U$ that are 
predominantly regular but all have a small chaotic admixture.  
The eigenstates $|\psich\rangle$ of $\Uch$ live in the chaotic 
region of $U$ and decay into the regular island
(Fig.~\ref{fig:quantum_maps}f). 

As $|\psireg\rangle$ and $|\psich\rangle$ are eigenstates of different operators
$\Ureg$ and $\Uch$, they are not necessarily orthogonal, 
$\langle\psich|\psireg\rangle=\chi$ with $0\leq|\chi|\ll 1$.
In order to apply Fermi's golden rule we introduce orthonormalized states
$\vert\widetilde{\psi}_{\text{reg}}\rangle = \vert\psi_{\text{reg}}\rangle$, 
$\vert\widetilde{\psi}_{\text{ch}}\rangle = (\vert\psi_{\text{ch}}
\rangle-\chi^{*}\vert\psi_{\text{reg}}\rangle)/\sqrt{1-|\chi|^2}$,
leading to 
$\langle\widetilde{\psi}_{\text{ch}}|\widetilde{\psi}_{\text{reg}}\rangle=0$. 
We find up to first order in $\chi$ for the coupling matrix element 
\begin{eqnarray}
  \label{eq:coupmatelement}
    v & \approx & \langle\psi_{\text{ch}}\vert U-
                      \Ureg\vert\psi_{\text{reg}}\rangle,
\end{eqnarray}
which can be inserted into 
Eq.~\eqref{eq:tunneling_rate_1}. The appearing term 
$\sum_{\text{ch}}|\psich\rangle\langle\psich|$ is semiclassically equal to
the projection operator onto the chaotic region. It can be approximated as
$\mathds{1}-P_{\text{reg}}$, where $P_{\text{reg}}$ is a projector onto the 
regular island. This yields
\begin{equation}
  \label{eq:tunneling_rate_final}
    \gamma_m \approx 
       \left\Vert (\mathds{1}-P_{\text{reg}})(U - \Ureg)
       |\psireg^m\rangle\right\Vert ^{2}
\end{equation}
as our main result, which involves properties of the fictitious regular 
system $\Ureg$ and the difference $U-\Ureg$. It allows for determining tunneling 
rates from the regular state on the $m$-th quantized torus 
to the chaotic sea. 

The most difficult step in applying Eq.~\eqref{eq:tunneling_rate_final} 
is the determination of the fictitious integrable system $\Ureg$, defined by 
a time-independent 1D Hamiltonian $\Hreg(p,q)$. On the one hand 
its dynamics over one time unit should resemble the classical motion 
corresponding to $U$ within the regular island as closely as possible.
As a result the contour lines of $\Hreg(p,q)$ in phase 
space (Fig.~\ref{fig:quantum_maps}b) approximate the KAM-curves of the 
classical map (Fig.~\ref{fig:quantum_maps}a).
On the other hand the function $\Hreg(p,q)$ should extrapolate sufficiently 
smoothly to the remaining phase space region. This is essential for the quantum 
eigenstates of $\Hreg$ to have reasonable tunneling tails in the neighborhood of 
the regular island. Finding an optimal $\Hreg$ is a difficult task. 
In fact, it will resemble the dynamics within the island with finite accuracy 
only, due to the generic presence of small resonance chains and the complicated 
structure of tori at the boundary of a regular island. Similar problems
appear for the analytic continuation of a regular torus into complex space due 
to the existence of a so-called natural boundary 
\cite{Wil1986,Cre1998,ShuIkeOniTakComb,BroSchUllEltComb}.
For the quantum tunneling problem at not too small $\heff$ and thus for a finite 
phase-space resolution, however, such an $\Hreg$ with limited accuracy can be 
good enough. We will discuss below two approaches \cite{Sch1988} leading to a 
sufficiently good $\Hreg$ for the prediction of tunneling rates.
Quantizing $\Hreg$ yields the required quantum mechanical operator 
$\Ureg=\ue^{-\ui\widehat{H}_{\text{reg}}/\hbareff}$ with corresponding
eigenfunctions $|\psireg^m\rangle$. For the numerical evaluation of  
Eq.~\eqref{eq:tunneling_rate_final} in 
Fig.~\ref{fig:distorted_island_amphib} it is convenient to replace 
$\Ureg|\psireg^m\rangle$ by $\ue^{-\ui E_m/\hbareff}|\psireg^m\rangle$ and
approximate $P_{\text{reg}}\approx\sum|\psireg^m\rangle\langle\psireg^m|$, 
where the sum extends over $m=0,\,1,\,\dots,\,\lfloor A/\heff-1/2\rfloor$.

\begin{figure}[t]
  \begin{center}
    \includegraphics[angle = 0, width = 85mm]
                     {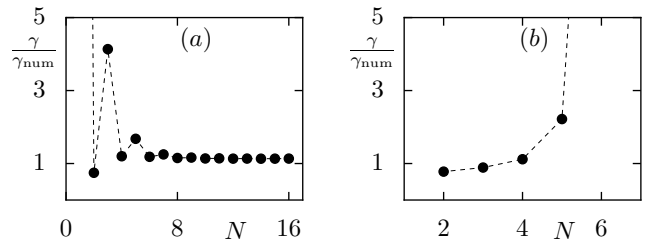}
    \vspace*{-0.4cm}
    \caption{Predicted tunneling rate, Eq.~\eqref{eq:tunneling_rate_final}, 
             normalized by the numerical value
             for $m=0$, $\heff=1/32$ vs order $N$ of $\Hreg$ corresponding to
             (a) Fig.~\ref{fig:distorted_island_amphib} ($N=10$) and 
             (b) Fig.~\ref{fig:rates_standard_map} ($N=4$).
          }
    \label{fig:convergence}
    \vspace*{-0.7cm}
  \end{center}
\end{figure}

In the following we will discuss the application of 
Eq.~\eqref{eq:tunneling_rate_final} for 1D kicked systems
$H(p,q,t) = T(p) + V(q) \sum_{n} \delta(t - n)$,
yielding the classical mapping: $q_{t+1}=q_{t}+T'(p_{t})$, 
$p_{t+1}=p_{t}-V'(q_{t+1})$. The corresponding quantum map over one kick 
period is $U = \exp[-\ui V(\widehat{q})/\hbareff]\exp[-\ui 
T(\widehat{p})/\hbareff] = \UV \UT$,
where $\hbareff$ is the ratio of Planck's constant $\hbar$ to the area of a 
phase-space unit cell. We consider a compact phase space with periodic 
boundary conditions for $q \in [-1/2, \, 1/2]$ and $p \in [-1/2, \, 1/2]$. 
In order to avoid the influence of resonances and cantori on the tunneling 
rates we use a system containing one regular island with very small resonance 
chains and a narrow transition region to a homogeneous chaotic sea. It is 
obtained by an appropriate choice of the functions $V'(q)$ and $T'(p)$ 
\cite{Map_amphib}. The phase space is shown 
in the Husimi function insets of Fig.~\ref{fig:distorted_island_amphib}. 
After determining $\Ureg$ and $|\psireg^m\rangle$ as described in the last 
paragraph we predict tunneling rates by evaluating 
Eq.~\eqref{eq:tunneling_rate_final}. 
Fig.~\ref{fig:distorted_island_amphib} 
shows a comparison to tunneling rates, 
determined numerically by absorbing boundary conditions at $q=\pm 1/2$ and 
taking twice the distance between the eigenvalue of the $m$-th regular state 
and the unit circle. We find excellent agreement for the tunneling 
rates $\gamma_m$ over $10$ orders of magnitude. 
The deviations for the smallest $\gamma$ can be attributed 
to the beginning of the resonance-assisted tunneling regime.
We determine $\Hreg$ using the Lie-transformation method \cite{LicLie1983}.
With increasing $N$, the tunneling rates following from 
Eq.~\eqref{eq:tunneling_rate_final} converge to a constant value 
(see Fig.~\ref{fig:convergence}a) and we choose $N=10$ for the
predictions in Fig.~\ref{fig:distorted_island_amphib}.
Note, that for sufficiently high $N$ (not shown in 
Fig.~\ref{fig:convergence}a) $\Hreg$ and the prediction 
for $\gamma$ are expected to diverge.

We now demonstrate that an analytical evaluation of 
Eq.~\eqref{eq:tunneling_rate_final} is possible for our example system.
We define functions $\widetilde{V}(q)$ and $\widetilde{T}(p)$ by a low
order Taylor expansion of $V(q)$ and $T(p)$, respectively, around the center 
of the regular island 
\cite{Fct_tilde}.
This results in a unitary operator $\UVreg\UTreg$ with the following
properties: 
(i) The corresponding classical dynamics is not necessarily regular.
(ii) It is close, however, to a regular quantum map $\Ureg$ beyond the border 
of the island and can therefore be used in Eq.~\eqref{eq:tunneling_rate_final} 
instead of $\Ureg$.
(iii) Within the island it has an almost identical classical dynamics as $U$. 
Therefore $(U-\UVreg\UTreg)|\psireg\rangle$ has almost all of its weight in 
the chaotic region and the projection operator $\mathds{1}-P_{\text{reg}}$ 
can be neglected in Eq.~\eqref{eq:tunneling_rate_final}.
With the definitions
$\mathds{1}+\epsV \equiv \ue^{-\frac{\ui}{\hbareff}\left[V(\widehat{q})
-\widetilde{V}(\widehat{q})\right]}$
and $\mathds{1}+\epsT \equiv \ue^{-\frac{\ui}{\hbareff}\left[T(\widehat{p})
-\widetilde{T}(\widehat{p})\right]}$ one obtains
$\gamma_m = \Vert\UVreg [\epsV + \epsT + \epsV \epsT ] \UTreg 
| \psireg^m \rangle\Vert^2$.
We find that typically the third contribution is negligible, leading to
\begin{eqnarray}
 \begin{split}
  \label{eq:tunneling_rate_2}
    \gamma_m \approx 
      2 \int \ud q \left\vert \psireg^m (q)  \right
      \vert^{2} \left[ 1-\cos \left( \frac{V(q)-\widetilde{V}(q)}{\hbareff}
        \right)\right] \\
      + 2 \int \ud p \left\vert \psireg^m (p)  \right
      \vert^{2} \left[ 1-\cos \left(\frac{T(p)-\widetilde{T}(p)}{\hbareff}
        \right)\right].
 \end{split}
\end{eqnarray}
In the last step the sums over the discrete position and momentum values have 
been replaced by integrals, which is valid in the semiclassical limit.
Agreement with the direct evaluation of Eq.~\eqref{eq:tunneling_rate_final} 
is found (not shown).
If an analytical WKB expression for the regular states $|\psireg^m\rangle$ is 
known, Eq.~\eqref{eq:tunneling_rate_2} can be evaluated further.
This is the case for a different parameter set
\cite{Map_amphib_2} which yields a tilted harmonic oscillator like island 
embedded in a chaotic sea.  
We approximate $V(q)-\widetilde{V}(q)$ and 
$T(p)-\widetilde{T}(p)$ by linear functions and use 
a WKB ansatz for the regular wave function. It turns out that the integral 
is proportional to the square of the modulus of the regular wave 
function at the border of the regular island. 
We obtain
\begin{eqnarray}
\label{eq:wkb_formula}
   \gamma_{m} =  
              c\,\frac{\heff}{\beta_m} \;
              \text{exp}\Bigg(-\displaystyle\frac{2A}{\heff}\left[\beta_m - 
              \alpha_m \ln \left(
              \frac{1+\beta_m}{\sqrt{\alpha_m}}\right)\right]\Bigg)
\end{eqnarray}
as the semiclassical prediction for the tunneling rate of the $m$-th regular 
state, where $\alpha_m=(m+1/2)(A/\heff)^{-1}$, $\beta_m = \sqrt{1-\alpha_m}$, 
$A \approx 0.28$ is the area of the regular island and $c\approx 1$ 
is $\heff$ independent by a rough semiclassical estimate.
The prediction, Eq.~\eqref{eq:wkb_formula}, gives excellent agreement 
with numerically determined data over $10$ orders of magnitude in $\gamma$ 
(not shown).
Let us make the following remarks concerning Eq.~\eqref{eq:wkb_formula}:
(i) The only information about this non-generic island with constant
rotation number is $A/\heff$ as in Ref.~\cite{PodNar2003}.
(ii) While the term in square brackets semiclassically approaches $1$, 
it is relevant for large $\heff$.
(iii) In contrast to Eq.~\eqref{eq:tunneling_rate_2}, where the chaotic
properties are contained in the differences $V(q)-\widetilde{V}(q)$ and
$T(p)-\widetilde{T}(p)$, they appear in the prefactor $c$ via the linear 
approximation of these differences.

The paradigmatic model of quantum chaos is the standard map ($T(p)=p^2/2$, 
$V(q)=-K/(4\pi^2)\cos(2\pi q)$), which for $K=2.9$ has a large generic 
regular island. 
Absorbing boundary conditions at $q=\pm 1/2$ lead to strong 
fluctuations of the numerically determined tunneling rates as a function of 
$\heff$, presumably due to 
cantori. When choosing $q=\pm 1/4$, which is closer to the island, we find 
smoothly decaying tunneling rates (dots in Fig.~\ref{fig:rates_standard_map}). 
Evaluating Eq.~\eqref{eq:tunneling_rate_final} gives good agreement with
these numerical data. Note that this is the first quantitative prediction 
of regular-to-chaotic tunneling rates for the standard map. 
Here we determine $\Hreg$ by first using the frequency map analysis \cite{LasFroCel1992}
for characterizing the properties of the regular island. This information is 
used to find the optimal 2D Fourier series of order $N$ for $\Hreg$.
The tunneling rates following from Eq.~\eqref{eq:tunneling_rate_final} show for 
increasing $N$ the expected divergence (see Fig.~\ref{fig:convergence}b). 
For the predictions in Fig.~\ref{fig:rates_standard_map} we choose $N=4$ as the 
largest order before this divergence.

We now want to discuss the relation of our approach to previous studies. 
The semiclassical formula presented in Ref.~\cite{PodNar2003} 
(dotted lines in Figs.~\ref{fig:distorted_island_amphib}, 
\ref{fig:rates_standard_map}) deviates from 
numerically determined tunneling rates.  
It works best for the case of constant 
rotation number, 
which according to Ref.~\cite{She2005} is the approximation used in 
Ref.~\cite{PodNar2003}. However, it seems to be not generally applicable. 
The system studied in Ref.~\cite{SheFisGuaReb2006} can be approximated by 
a 1D Hamiltonian $\Hreg(p,q)$ with a cubic potential. 
Here the tunneling path ends far away from the island.
In such a case our result is also applicable but the use of the WKB expression 
presented in Ref.~\cite{SheFisGuaReb2006} is more convenient.
In general situations, however, the main contribution comes 
from tunneling to the neighborhood of the regular island as seen e.g.\ from 
Eq.~\eqref{eq:tunneling_rate_2}. 
 We also performed successful tests on the tunneling system 
investigated in Ref.~\cite{IshTanShu2006}.

In summary, we have derived a quantum mechanical formula 
Eq.~\eqref{eq:tunneling_rate_final} for the tunneling rates, which involves the 
fictitious integrable system $\Ureg$ and the difference $U-\Ureg$. It is the 
basis for deriving semiclassical expressions, which we demonstrated with 
Eqs.~\eqref{eq:tunneling_rate_2} and \eqref{eq:wkb_formula} for the case of 
a fictitious regular system, that is well approximated by a kicked system.
Still there are open questions about dynamical tunneling from a regular
island to the chaotic sea:
(i) Which properties of the regular island 
(e.g. size, winding number, shape) and which properties of the 
chaotic sea are relevant in general? (ii) Can the approach be combined with the 
resonance-assisted tunneling description and how can cantori be accounted for? 
(iii) How can it be generalized to time-independent Hamiltonian systems, 
in particular billiards?
We hope that our approach with a fictitious integrable system will allow to
answer these questions.

We thank S.~Fishman, P.~Schlagheck, A.~Shudo, and S.~Tomsovic for useful 
discussions, and the DFG for financial support.


\begin{thebibliography}{10}

\bibitem{WKBInstcomb}
E.~Merzbacher, {\em Quantum mechanics\/} (Wiley, New York, 1998);
E.~Gildener and A.~Patrascioiu, Phys.~Rev.~D {\bf 16}, 423 (1977).

\bibitem{DavHel1981}
M.~J.~Davis and E.~J.~Heller, J.~Chem.~Phys. {\bf 75}, 246 (1981).

\bibitem{HanOttAnt1984}
J.~D.~Hanson, E.~Ott, and T.~M.~Antonsen, Phys.~Rev.~A {\bf 29}, 819 (1984).

\bibitem{Wil1986}
M.~Wilkinson, Physica D {\bf 21}, 341 (1986).

\bibitem{BohTomUll1993}
O.~Bohigas, S.~Tomsovic, and D.~Ullmo, Phys.~Rep. {\bf 223}, 43 (1993).

\bibitem{TomUll1994}
S.~Tomsovic and D.~Ullmo, Phys.~Rev.~E {\bf 50}, 145 (1994).

\bibitem{Tom1998}
S.~Tomsovic, J.~Phys.~A {\bf 31}, 9469 (1998).

\bibitem{Cre1998}
S.~C.~Creagh, in {\em Tunneling in complex systems\/} 
(World Scientific, Singapore, 1998).

\bibitem{FriDor1998}
S.~D.~Frischat and E.~Doron, Phys.~Rev.~E {\bf 57}, 1421 (1998). 

\bibitem{ShuIkeOniTakComb}
A.~Shudo and K.~S.~Ikeda, Phys.~Rev.~Lett. {\bf 74}, 682 (1995); 
Physica D {\bf 115}, 234 (1998); 
T.~Onishi, A.~Shudo, K.~S.~Ikeda, and K.~Takahashi, Phys.~Rev.~E {\bf 64}, 
025201(R) (2001).

\bibitem{BroSchUllEltComb}
O.~Brodier, P.~Schlagheck, and D.~Ullmo, Phys.~Rev.~Lett. {\bf 87}, 064101 (2001); 
Ann.~Phys. {\bf 300}, 88 (2002); 
C.~Eltschka and P.~Schlagheck, Phys.~Rev.~Lett. {\bf 94}, 014101 (2005).

\bibitem{WimSchEltBuc2006}
S.~Wimberger, P.~Schlagheck, C.~Eltschka, and A.~Buchleitner, 
Phys.~Rev.~Lett. {\bf 97}, 043001 (2006).
  
\bibitem{SheFisGuaReb2006}
M.~Sheinman, S.~Fishman, I.~Guarneri, and L.~Rebuzzini, Phys.~Rev.~A {\bf 73}, 
052110 (2006).

\bibitem{PodNar2003}
V.~A.~Podolskiy and E.~E.~Narimanov, Phys.~Rev.~Lett. {\bf 91}, 263601 (2003); 
see Ref.~\cite{She2005} for a corrected formula, which is used in the figures.

\bibitem{She2005}  
M.~Sheinman, Masters thesis, Technion, 2005, Eq.~(A.14).

\bibitem{SteOskRai2001}
D.~A.~Steck, W.~H.~Oskay, and M.~G.~Raizen, Science {\bf 293}, 274 (2001).
  
\bibitem{HenHafBroHecHelMcKMilPhiRolRubUpc2001}
W.~K.~Hensinger et al., Nature {\bf 412}, 52 (2001).

\bibitem{FroWilHayEavSheMIuHen2002}
T.~M.~Fromhold et al., Phys.~Rev.~B {\bf 65}, 155312 (2002).

\bibitem{SchOttKetDit2001}
H.~Schanz, M.-F.~Otto, R.~Ketzmerick, and T.~Dittrich, Phys.~Rev.~Lett.
{\bf 87}, 070601 (2001).

\bibitem{BaeKetMon200507}
A.~B\"acker, R.~Ketzmerick, and A.~G.~Monastra, 
Phys.~Rev.~Lett. {\bf 94}, 054102 (2005); 
Phys.~Rev.~E {\bf 75}, 066204 (2007).

\bibitem{WieHen2006}
J.~Wiersig and M.~Hentschel, Phys.~Rev.~A {\bf 73}, 031802(R) (2006).

\bibitem{BerRob1984}
M.~V.~Berry and M.~Robnik, J.~Phys.~A {\bf 17}, 2413 (1984).

\bibitem{FeiBaeKetRotHucBur2006}
J.~Feist, A.~B\"acker, R.~Ketzmerick, S.~Rotter, B.~Huckestein, and 
J.~Burgd\"orfer, Phys.~Rev.~Lett. {\bf 97}, 116804 (2006).

\bibitem{BerBalTabVor1979}
M.~V.~Berry, N.~L.~Balazs, M.~Tabor, and A.~Voros, 
Ann.~Phys. {\bf 122}, 26 (1979).

\bibitem{Rel_idea} A related idea was presented in \cite{BohTomUll1993}.

\bibitem{Sch1988}
For a different approach see R.~Scharf, J.~Phys.~A {\bf 21}, 4133 (1988) 
and references therein.

\bibitem{Map_amphib} We start with functions $t'(p)=1/2 \pm (1-2p)$ for
  $0 < \pm p < 1/2$ and $v'(q)=-rq+Rq^2$ for $-1/2 < q < 1/2$. 
  Smoothing the periodically extended functions with a Gaussian, 
  $G(z)=\exp(-z^2/2 \varepsilon^2)/\sqrt{2\pi\varepsilon^2}$, gives analytic 
  functions $T'(p)=\int \ud z \, t'(p+z)G(z)$ and 
  $V'(q)=\int \ud z \, v'(q+z)G(z)$. We take 
  $r=0.26$, $R=0.4$, and $\varepsilon=0.005$.
  A similar model, restricted to constant rotation number ($R=0$), 
  has been used in \cite{SchOttKetDit2001,BaeKetMon200507}.

\bibitem{LicLie1983}
A.~J.~Lichtenberg and M.~A.~Liebermann, {\em Regular and Stochastic Motion\/}
  (Springer-Verlag, New York, 1983).

\bibitem{Fct_tilde} For the map defined in \cite{Map_amphib} we get 
                    $\widetilde{T}'(p)=3/2-2p$ and 
                    $\widetilde{V}'(q)=-rq+R(q^2+\varepsilon^2)$.

\bibitem{Map_amphib_2} We use the map defined in \cite{Map_amphib} 
         with parameters $r=0.65$, $R=0$, and $\varepsilon=0.002$.

\bibitem{LasFroCel1992}
J.~Laskar, C.~Froeschl\'e, and A.~Celletti,
Physica D {\bf 56}, 253 (1992).

\bibitem{IshTanShu2006}
A.~Ishikawa, A.~Tanaka, and A.~Shudo, J.~Phys.~A {\bf 40} F397 (2007).

\end{thebibliography}
\end{document}